\pdfoutput=1
\documentclass[conference]{IEEEtran}

\usepackage{graphicx}
\usepackage{color}
\usepackage{placeins}
\usepackage{float}
\usepackage{tabularx,colortbl}
\usepackage{amssymb}
\usepackage{amsthm}
\usepackage{cite}
\usepackage{amsmath}
\usepackage{algorithm}
\usepackage{algorithmic}
\usepackage[a4paper, bottom=0.95in]{geometry}
\usepackage{marvosym}

\usepackage{caption2}
\usepackage{geometry}
\theoremstyle{plain}
\newtheorem{thm}{Theorem}

\theoremstyle{plain}

\IEEEoverridecommandlockouts

\begin{document}

\title{Movable Cell-Free Massive MIMO For High-Speed Train Communications: A PPO-Based Antenna Position Optimization}
\author{
    \IEEEauthorblockN{Jie Dai\IEEEauthorrefmark{1}, 
    Yuchen Liu\IEEEauthorrefmark{1}, 
    Jiakang Zheng\IEEEauthorrefmark{1}, 
    Ruichen Zhang\textsuperscript{$\diamond$}, 
    Jiayi Zhang\IEEEauthorrefmark{1}, and 
    Bo Ai\IEEEauthorrefmark{1}}
    \IEEEauthorblockA{\IEEEauthorrefmark{1}School of Electronic and Information Engineering, Beijing Jiaotong University, Beijing 100044, China}
    \IEEEauthorblockA{\textsuperscript{$\diamond$}College of Computing and Data Science, Nanyang Technological University, Singapore}
    \IEEEauthorblockA{E-mail: \{22211221, 22211232, jiakangzheng, zhangjiayi, boai\}@bjtu.edu.cn; ruichen.zhang@ntu.edu.sg}
\thanks{This work was supported in part by the China Postdoctoral Science Foundation under Grant Number 2024M760195; in part by the Talent Fund of Beijing Jiaotong University 2024XKRC085; in part by ZTE Industry-University-Institute Cooperation Funds under Grant No. IA20240709018; in part by Jiangxi Province Science and Technology development Programme under Grants 20242BCC32016.}
}
\maketitle
\vspace{-1cm}
\begin{abstract}
In recent years, high-speed trains (HSTs) communications have developed rapidly to enhance the stability of train operations and improve passenger connectivity experiences. However, as the train continues to accelerate, urgent technological innovations are needed to overcome challenges such as frequency handover and significant Doppler effects. In this paper, we present a novel architecture featuring movable antennas (MAs) to fully exploit macro spatial diversity, enabling a cell-free (CF) massive multiple-input multiple-output (MIMO) system that supports high-speed train communications. Considering the high likelihood of line-of-sight (LoS) transmission in HST scenario, we derive the uplink spectral efficiency (SE) expression for the movable CF massive MIMO system. Moreover, an optimization problem is formulated to maximize the sum SE of the considered system by optimizing the positions of the antennas. Since the formulated problem is non-convex and highly non-linear, we improve a deep reinforcement learning algorithm to address it by using proximal policy optimization (PPO). Different from traditional optimization approaches, which optimize variables separately and alternately, our improved PPO-based approach optimizes all the variables in unison. Simulation results demonstrate that movable CF massive MIMO effectively suppresses the negative impact of the Doppler effect in HST communications.
\end{abstract}

\IEEEpeerreviewmaketitle

\section{Introduction}
 Recently, HSTs have developed rapidly, becoming a vital component of the transport infrastructure. However, ensuring high-quality communication services for passengers in HSTs remains an important issue. A key characteristic of HSTs is the high likelihood of line-of-sight (LoS) link, which minimizes the need to consider scattering and reflection effects. This is mainly due to the LoS-favorable design of HST systems \cite{9690475}. In addition, HST wireless communication systems need to overcome challenges caused by high speed, such as frequency handover and Doppler frequency offset (DFO) \cite{zheng2022cell}. 

To address these challenges, one promising solution is to adopt a distributed architecture called cell-free (CF) massive multiple-input multiple-output (MIMO)\cite{10556753}. CF massive MIMO is a distributed MIMO system that involves multiple access points (APs) working together to serve a relatively smaller number of user equipments (UEs)\cite{ngo2017cell}. APs are connected to a central processing unit (CPU) via a backhaul network and coherently serve all UEs by spatial multiplexing over the same time-frequency resource. The system can achieve high and uniform spectral efficiency (SE) within wireless networks, outperforming both small cell and massive MIMO configurations \cite{zhang2020prospective,bjornson2019making,zheng2024rate}. The presence of numerous distributed APs can effectively mitigate the issue of frequency handover between adjacent APs. This configuration facilitates seamless transitions and improves communication reliability, thereby enhancing user experience in high-speed environments.

In addition, in order to decrease the effect of DFO in HST, we find that movable antenna (MA) is possible solution\cite{zheng2024flexible}.\footnote{In the case of high-speed communications, although the antenna position needs to be reconfigured in a few milliseconds or less to be effective, this is feasible for mobile antennas with the help of micro-electromechanical systems (MEMS) due to the small scale of movement.} MAs technologies have conventionally been explored within the field of antenna design and recently gained significant focus in the field of wireless communications\cite{zheng2024mobile}. This interest stems from their capability for dynamical positioning of transmit and receive antennas within a specified area \cite{zhu2023movable_opti}. Unlike traditional fixed-position antennas (FPAs), which may lead to substantial fading at particular locations during specific time and frequency resource blocks, FAs/MAs offer the benefit of flexible mobility for antennas\cite{mei2024movable}. This adaptability circumvents deep-fading zones, effectively reshaping the wireless channel to enhance communication performance. Therefore, MAs may be particularly effective in mitigating performance degradation caused by Doppler frequency shifts, and improving signal reliability in high-speed scenarios.

While MA holds significant promise, it also introduces new challenges, specifically in determining of the optimal positioning of the received signal to maximize SE. To overcome this challenge, various optimization methods, such as a graph-based approach \cite{mei2024movable}, particle swarm optimization (PSO) \cite{pi2023multiuser}, and a method using zero force (ZF) and minimum mean square error (MMSE) \cite{zhu2023movable_opti}, have been proposed to find the suitable antenna positions. Optimal position selection is straightforward for a single FA/MA, while the position selection becomes significantly more complex and difficult to solve optimally in the case of multiple FAs/MAs. Therefore, we improve a deep reinforcement learning (DRL) algorithm to address this complex and non-convexity problem by using proximal policy optimization (PPO)\cite{zhang2023energy}. Unlike traditional
optimization approaches, which optimize variables separately and alternately, our improved PPO-based approach optimizes all the variables in unison. To evaluate the performance of the PPO-based algorithm, it is compared with the greedy algorithm and the random search algorithm. Simulation results indicate that the performance of the PPO-based algorithm outperforms than the other two algorithms. In this paper, the specific technical contributions are as follows:
\begin{enumerate}
  \item We propose a movable CF massive MIMO system for HST communications and investigate the impact of DFO on its SE performance.
   \item  We improve the PPO algorithm to optimize the position of the mobile antenna, which can effectively suppress the impact of DFO. 
\end{enumerate}


\section{System Model}\label{se:model}
\begin{figure}[t]
\vspace{0.2cm}
\centering
\setlength{\abovecaptionskip}{0.cm}
\includegraphics[scale=0.55]{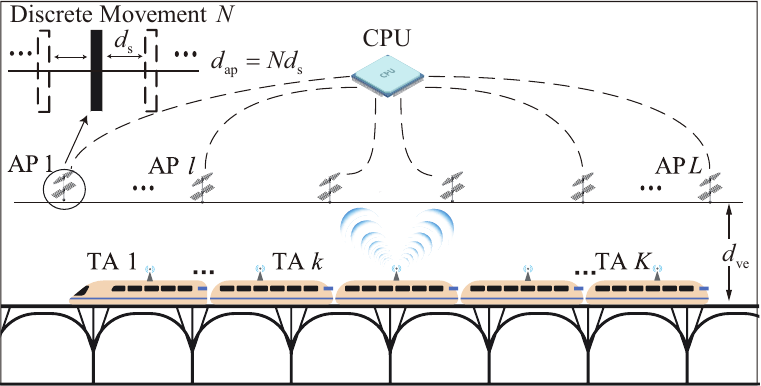}
\caption{HST CF massive MIMO systems with MA.} \vspace{-4mm}
\label{HST}
\end{figure}

As illustrated in Fig.~\ref{HST}, we consider a CF massive MIMO system with $K$ single-antenna train and $L$ single-antenna APs. The positions of the antennas on the trains are assumed to remain fixed, while the APs are equipped with MA. APs are connected to a central processing unit (CPU) through a backhaul network, and they serve all train antennas (TAs)  over the same time-frequency resources \cite{zheng2023asynchronous}. The system architecture is defined in a two-dimensional (2D) Cartesian coordinate. Let the \(l\)-th AP be denoted as \({r_l} = [{x_{l}},{d_{{\text{ve}}}}]\), \(l = 1, \ldots ,L\), where \(x_l\) is the horizontal axis value and \({d_\text{ve}}\) represents the constant vertical distance separating the APs from the TAs. The position of the $k$-th TA can be represented by the coordinates \({t_k} = [{x_{k}},0]\). Thus, the distance along the horizontal axis separating the $l$-th AP from the $k$-th TA is defined by \(d_{kl}^{\text{ho}} = \left| {{x_{l}} - {x_{k}}} \right|\), where the straight line distance is given by
\begin{align}
{d_{kl}} = \sqrt {{{(d_{kl}^{\text{ho}})}^2} + {{({d_{\text{ve}}})}^2}}.
\end{align}
Moreover, the angle of arrival (AoA) between $l$-th AP and $k$-th TA is \(\theta _{kl}\), and its cosine is given by
\(\cos ({\theta _{kl}}) = {{d_{kl}^{\text{ho}}}}/{{{d_{kl}}}}\).

\subsection{Channel Model}
Given the significant LoS component in HST communications, the channel coefficient between $l$-th AP and $k$-th TA is determined as \({\beta _{kl}} = {\beta _0}{({d_{kl}})^{ - \alpha }}\), where \(\alpha  = 3\) is the path loss exponent, and \({\beta _0} = {10^{ - 12}}\) denotes the path loss per km between TA and AP \cite{zheng2022cell}. Then the channel coefficient between $l$-th AP and $k$-th TA is determined as
\begin{align}\label{h_kl}
{h_{kl}} = \sqrt {{\beta _{kl}}} \exp \left( {j\frac{{2\pi }}{\lambda }{d_{kl}}} \right).
\end{align}

The DFO due to the train's fast movement is then considered, and the channel gain is represented as:
\begin{align}
{g_{kl}} = \sqrt {{\beta _{kl}}} \exp \left( {j\frac{{2\pi }}{\lambda }\left( {{d_{kl}} + w\cos \left( {{\theta _{kl}}} \right)} \right)} \right),
\end{align}
where \(w = \frac{{fvT}}{c}\) is the maximum normalized DFO, and $T$ represents the duration of signal sampling. Morever, $v$, $c$, and $f$ represent train speed, light speed in vacuum, and carrier frequency, respectively.

For considered MA model, it is assumed that the AP antennas are adjustable along a uniform linear array (ULA) parallel to the track with the help of drive components\cite{wang2024generative}. Specifically, the ULA has $N$ discrete positions and the total length is \({d_{{\rm{ap}}}} = N{d_{\rm{s}}}\), where \({d_{\rm{s}}}\) represents the minimum distance that the MA can move. Thus, for the signal of the $k$-th TA, The phase shift caused by the position modification of the $l$-th AP antenna can be represented as
\begin{equation}
{f_{kl}} = j\frac{{2\pi }}{\lambda }{n_l}{d_\text{s}}\cos \left( {{\theta _{kl}}} \right),
\end{equation}
where \({n_{l}}\) represents the position of the antenna on $l$-th AP. Note that due to the relatively small order of magnitude of the antenna spacing compared to the straight line distance \(d_{{kl}}\), AoA is assumed to remain unchanged as the antenna is repositioned\cite{olyaee2024user}. 
\subsection{Uplink Transmission}
Considering the phase shift due to antenna movement, the total signal received by $l$-th AP is expressed as\cite{wang2024optimal}
\begin{align}
{y_l} = \sum\limits_{i = 1}^K {\sqrt {{p_k}} {g_{il}}\exp \left( {j\frac{{2\pi }}{\lambda }{n_l}{d_{\rm{s}}}\cos \left( {{\theta _{il}}} \right)} \right)} {x_k} + {w_l},
\end{align}
 where \({w_l}\) is the complex circular white Gaussian noise, while ${p_k}$ represents the uplink transmission power of $k$-th TA . 

To detect the signal from $k$-th TA, $l$-th AP multiplies the conjugate of the channel estimate (i.e., perfect channel state information ) as described in \eqref{h_kl}. The resulting quantity \({{\mathord{\buildrel{\lower3pt\hbox{$\scriptscriptstyle\smile$}} 
\over y} }_{kl}} = h_{kl}^ * {y_l}\) is then transmitted to the CPU via the backhaul network. The CPU subsequently obtains \({{{\hat y}}_k}\), as given in \eqref{y}, on the next page, where \({{\rm{D}}{{\rm{S}}_k}}\) denotes the desired signal, and \({{\rm{U}}{{\rm{I}}_i}}\) represents the interference caused by transmitted data from other TAs.
\begin{figure*}[t!]
\normalsize
\begin{align}\label{y}
{{\hat y}_k}\!\! =\!\! \underbrace {\sum\limits_{l = 1}^L {\sqrt {{p_k}} h_{kl}^ * {h_{kl}}\exp \left( {\!j\frac{{2\pi }}{\lambda }\!\cos \left( {{\theta _{kl}}} \right)\left( {w\! + \!{n_l}{d_{\rm{s}}}} \right)} \!\right)} }_{{\rm{D}}{{\rm{S}}_k}}\!{x_k}\! +\! \sum\limits_{i \ne k}^K {\underbrace {\sum\limits_{l = 1}^L {\sqrt {{p_i}} h_{kl}^ * {h_{il}}\exp \left( {\!j\frac{{2\pi }}{\lambda }\!\cos \left( {{\theta _{il}}} \right)\left( {w \!+ \!{n_l}{d_{\rm{s}}}} \right)} \!\right)} }_{{\rm{U}}{{\rm{I}}_k}}}\!{x_i} \!+\! \sum\limits_{l = 1}^L {h_{kl}^ * {w_l}}.
\end{align}
\hrulefill
\end{figure*}

\begin{figure*}[t!]
\normalsize
\begin{align}\label{SE_k}
{\rm{SIN}}{{\rm{R}}_k} = \frac{{{{\left| {\sum\limits_{l = 1}^L {{p_k}{\beta _{kl}}\exp \left( {j\frac{{2\pi }}{\lambda }\left( {w\cos \left( {{\theta _{kl}}} \right) + {n_l}{d_{\rm{s}}}\cos \left( {{\theta _{kl}}} \right)} \right)} \right)} } \right|}^2}}}{{\sum\limits_{i \ne k}^K {{{\left| {\sum\limits_{l = 1}^L {{p_i}\sqrt {{\beta _{il}}{\beta _{kl}}} \eta ({d_{il}},{d_{kl}})\exp \left( {j\frac{{2\pi }}{\lambda }\left( {w\cos \left( {{\theta _{kl}}} \right) + {n_l}{d_{\rm{s}}}\cos \left( {{\theta _{kl}}} \right)} \right)} \right)} } \right|}^2} + \sum\limits_{l = 1}^L {{\beta _{kl}}{\sigma ^2}} } }}.
\end{align}
\hrulefill
\end{figure*}

\begin{thm}\label{thm1}
With the help of \eqref{y} and \cite{bjornson2017massive}, the SE of $k$-th TA is
\begin{align}\label{1}
{\rm{S}}{{\rm{E}}_k} = {\log _2}\left( {1 + {\rm{SIN}}{{\rm{R}}_k}} \right),
\end{align}
with ${\rm{SIN}}{{\rm{R}}_k}$ is given by
\begin{align}\label{2}
{\rm{SIN}}{{\rm{R}}_k} = \frac{{{{\left| {{\rm{D}}{{\rm{S}}_k}} \right|}^2}}}{{\sum\limits_{i \ne k}^K {{{\left| {\rm{U}{\rm{I}_i}} \right|}^2}}  + {\sigma ^2}\sum\limits_{l = 1}^L {{{\left| {{h_{kl}}} \right|}^2}} }}.
\end{align}
Morever, using \eqref{h_kl}, we can derive
\begin{align}
\label{3}&h_{kl}^ * {h_{kl}} = {\beta _{kl}},\\
h_{kl}^ * {h_{il}} = &\sqrt {{\beta _{kl}}{\beta _{il}}} \exp \left( {j\frac{{2\pi }}{\lambda }\left( {{d_{il}} - \label{4}{d_{kl}}} \right)} \right)\notag\\
 = &\sqrt {{\beta _{kl}}{\beta _{il}}} \eta \left( {{d_{il}},{d_{kl}}} \right).
\end{align}
Finally, substituting \eqref{3} and \eqref{4} into \eqref{2}, we can derive the expression for the SINR of the $k$-th TA as \eqref{SE_k} shown at the top of this page, thus completing the proof.
\end{thm}

In order to improve the overall performance of the system, our objective is to maximize the sum of SE across all TAs while optimizing AP antenna placement, the corresponding optimization problem is mathematically formulated as
\begin{subequations}
\begin{align} 
{P_0}:\mathop {\max }\limits_{{C_N}} \sum\limits_{k = 1}^K {\text{S}{\text{E}_k}} \\
{\text{s.t.}} \forall {n_l} \in [1:N],\\
{n_l} \in {C_N}.
\end{align}
\end{subequations}

\section{PPO-Based Algorithm}
Problem \({{\rm{P}}_0}\) is a multi-variable optimization problem with dynamic environments, which motivates us to develop an advanced DRL-based method. In this section, we first reformulate the Problem \({{\rm{P}}_0}\) as a Markov Decision Process (MDP). Then, we propose a PPO-based DRL algorithm to enhance communication performance in HST environments \cite{zhao2023ppo}.
\subsection{MDP Formulation}
An MDP provides a framework for decision-making in dynamic environments, consisting of an agent and a four-tuple \( < {\bf{s}},{\bf{a}},r,\gamma  > \), where the four elements represent the state space, action space, reward function, and discount factor. Accordingly, the action space, the state space, and reward are defined as follows.
\begin{itemize}
    \item \textbf{Action space:} The action space consists of the coordinates of the MAs on each AP. The position of the MA on the \(l\)-th AP is denoted by \(a_l\), and the action space represents the positions
of all MAs. Formally, the action space is defined as
    \begin{equation}
     {\bf{a}} = \left[ {{a_1}, \ldots ,{a_L}} \right], 
    \end{equation}
    where the cardinal number of the action space is \(L\).
\end{itemize}

\begin{itemize}
    \item \textbf{State space:} The state space comprises the selected action vector \(\bf{a}\), the instant reward \(r\), and the SE value for each TA. The selected action vector \(\bf{a}\) represents positions selected in the previous time step by the PPO-based algorithm. The instant reward \textit{r }is computed based on the current state and action, which is given by
    \begin{equation}
        r = \sum\limits_{i = 1}^K {\text{SE}_i}.
    \end{equation}
The state of all TAs \(\bf{u}\) is defined as
    \begin{equation}
        {\bf{u}} = \left[ {{\rm{S}}{{\rm{E}}_1}, \ldots ,{\rm{S}}{{\rm{E}}_K}} \right].
    \end{equation}
Thus, the state space is defined as
\begin{equation}
    \mathbf{s} = \left\{\mathbf{a}, r, \mathbf{u}\right\},
\end{equation}
where the cardinal number of the state space is \(\left( {L + 1 + K} \right)\).
\end{itemize}

\begin{itemize}
    \item \textbf{Reward:} In line with the optimization objective of problem \({{\rm{P}}_0}\), the reward function is defined as the sum SE across all TAs, which corresponds to \(r\) in the state space. It is subject to several key constraints, including AP converge area and power consumption. The objective is to maximize the reward function to enhance the communication performance in HST environments. 
\end{itemize}

\subsection{PPO Algorithm}
We develop a PPO-based DRL algorithm to solve the formulated MDP and learn the optimal policy. The PPO algorithm is one of the leading DRL algorithms within the class of policy gradient algorithms \cite{10480579}. It mitigates instability issues by using a clipped surrogate objective, which ensures that updates to the policy network are controlled.
\begin{algorithm}[t]
\caption{The PPO-Based Algorithm}
\begin{algorithmic}[1]
\STATE \textbf{Input:} Corresponding channels, current state ${\bf{s}}$ and reward \(r\);
\STATE \textbf{Output:} Action ${\bf{a}} = \left[ {{a_1}, \ldots ,{a_L}} \right]$;
\STATE \textbf{Initialization:} Initialize actor parameters $\theta_A$, critic parameters $\theta_C$ and clipping parameter \(\varepsilon\);
\FOR{episode = 1, 2, \dots, max episode}
    \STATE Initialize state $s_t$;
    \FOR{time step $t = 1$ to max time step}
        \STATE Choose action $a_t$ according to policy $\pi$;
        \STATE Perform action $a_t$, observe reward $r_t$ and next state $s_{t+1}$, store transition in the experience pool;
    \ENDFOR
    \STATE Sample transitions from experience pool;
    \STATE Calculate advantage estimate $A_t\left({\theta_A^{\text{old}}}\right)$ using \(n\)-\(step\) \(returns\);
    \STATE Compute the surrogate objective function $J\left(\theta_A\right)$;
    \STATE Compute the clipped surrogate objective function $J^{\text{CLIP}}\left(\theta_A\right)$;
    \STATE Perform gradient ascent on $J^{\text{CLIP}}\left(\theta_A\right)$ to update actor parameters $\theta_A$;
    \STATE Compute target state-value function $V_{\text{tar}}\left(s_t\right)$ using \(n\)-\(step\) \(returns\);
    \STATE Perform gradient descent on the mean square error to update critic parameters $\theta_C$;
\ENDFOR
\end{algorithmic}
\end{algorithm}

The parameters of the actor network, denoted as \(\theta_A\), are first initialized  randomly. The policy \(\pi\) is responsible for action selection.  The critic network estimates the state-value function with parameters \(\theta_C\), denoted by \(V\left(s_{t}\mid {\theta_{C}}\right)\). To simplify the optimization process, a surrogate objective function is employed, i.e.,
\begin{equation}
J(\theta_A) = \mathbb{E}_t \left[ \rho_t\left(\theta_A\right) A_t\left({\theta_A^{\text{old}}}\right) \right],
\end{equation}
where \(\theta_{A}^{\text{old}}\) denote the parameters of the actor network in the last episode. Here, \(\rho_t\left( \theta_A \right) \overset{\Delta}{=} \pi\left( a_t, s_t \mid {\theta_A} \right) \mathbin{/} \pi\left( a_t, s_t \mid {\theta_A^{\text{old}}} \right)\) represents the probability ratio of the current policy to the old one. The term \(A_t\left({\theta_A^{\text{old}}}\right)\) represents the advantage function, measuring the relative value of an action compared to the average action at a given state, which is calculated by \(n\)-\(step\) \(returns\), i.e.,
\begin{equation}
\begin{aligned}
A_t\left({\theta_A^{\text{old}}}\right) = & \, r(s_t, a_t) + \cdots + \gamma^{n+1}  V\left(s_{t+n+1} \mid {\theta_{C}^{\text{old}}}\right) \\
& - V\left(s_{t} \mid {\theta_{C}^{\text{old}}}\right),
\end{aligned}
\end{equation}
with \(\gamma  \in \left[ {0,1} \right)\) and \(\theta_{C}^{\text{old}}\) denoting the discount factor and the parameters of the critic network, respectively.

To address the issues of high variance and instability often encountered in traditional DRL methods, PPO algorithm introduces a clipping mechanism to the surrogate objective function, which is formulated as
\begin{equation}
\begin{split}
    J^{\text{CLIP}}\left(\theta_A \right) = &{\mathbb{E}_t }\left[\min \left(\rho_t(\theta_A) A_t\left({\theta_A^{\text{old}}}\right), \right.\right.\\
    &\left.\left. \text{clip}\left(\rho_t(\theta_A),1 - \varepsilon ,1 + \varepsilon \right)A_t\left({\theta_A^{\text{old}}}\right)\right)\right],
\end{split}
\end{equation}
where the \({\text{clip}}\left(x,1 - \varepsilon ,1 + \varepsilon \right)\) restricts \(x\) within the range \(\left[1 - \varepsilon ,1 + \varepsilon \right]\) and \(\varepsilon \) is a hyper-parameter controlling the clipping range. This clipping mechanism ensures that policy updates remain controlled, reducing the risk of divergence and enhancing training stability.

In addition, PPO algorithm is a policy-gradient approach, that updates \(\theta_A\) with the gradient of \(J^{\text{CLIP}}(\theta_A)\). The update rule for \(\theta_A\) is given by
\begin{equation}
\theta_A = \theta_A^{\text{old}} - \alpha_A \frac{1}{B} \sum_{t=1}^{B} \left( \nabla_{\theta_A} \tilde{J}^{\text{CLIP}}(\theta_A) \right),
\end{equation}
where \(\alpha_A\) represents the learning rate, and \(\tilde{J}^{\text{CLIP}}(\theta_A)\) indicates the specific instance of \(J^{\text{CLIP}}(\theta_A)\) at the \(t\)-th time step. Particularly, the mini-batch stochastic gradient descent (SGD) method is utilized to update the parameters, where samples \(\{s_t, a_t, r_t, s_{t+1}\}\) are selected randomly from the experience pool. In addition, the parameters \(\theta_C\) are updated using the mini-batch SGD method, with the mean squared error employed as the loss function, which is denoted as
\begin{equation}
\theta_C = \theta_{C}^{\text{old}} - \alpha_B \frac{1}{B} \sum_{t=1}^{B} \nabla_{\theta_A} \left( V\left(s_{t}\mid {\theta_C}\right) - V_{\text{tar}}(s_t) \right)^2,
\end{equation}
where \(\alpha_B\) represents the learning rate, and \(V_{\text{tar}}(s)\) denotes the target state-value function, which is calculated based on \(n\)-\(step\) \(returns\), i.e.,
\begin{equation}
V_{\text{tar}}(s_t) = r(s_t, a_t) + \cdots + \gamma^{n+1} V\left(s_{t+n+1}\mid {\theta_{C}^{\text{old}}}\right). 
\end{equation}

Note that in the whole process, the interactions are stored in an experience pool, which is used to sample data for training. The advantage function \(A_t\left({\theta_A^{\text{old}}}\right)\) is then calculated to measure the relative value of an action. Using the sampled data, the policy parameters \(\theta_A\) and \(\theta_C\) are updated through gradient ascent on the surrogate objective function. Therefore, PPO is efficient in adapting to minor system changes without retraining. The computational complexity of the PPO algorithm is mainly determined by the size of the DNNs, with complexity about \(O\left(\prod_{p=1}^{P} n_{p-1} \cdot n_p\right)\), where \(n_p\) is the number of neurons in the \(p\)-th layer.

The PPO algorithm optimizes the policy to maximize cumulative rewards, thereby refining the MA coordinates. Compared to existing schemes, the proposed algorithm incorporates a clipping mechanism, ensuring stable training and  improving the optimization process in HST environments. For clarity, the details of the proposed PPO-based algorithm are summarized in Algorithm 1.

\section{SIMULATION RESULTS}
In this section, we conduct numerical experiments to evaluate our PPO-based algorithm's effectiveness in optimizing the coordinates of MAs for HST communication with CF massive MIMO systems.
\subsection{Environment Setup}
The simulations consider a system with \(L\) APs uniformly distributed along a \(1000\) m high-speed railway and \(K\) TAs uniformly distributed on a \(300\) m HST. Each AP can position its MA at one of \(N\) discrete locations, equally spaced at fixed intervals. The carrier frequency is set to \(fc = 1.2\) GHz, with a signal sampling duration of \(T = 0.4\) ms \cite{9099874}. The signals transmitted by TA have a bandwidth of \(B = 20\) MHz and a noise power of \({\sigma ^2} =  - 96\) dBm. 

The PPO-based algorithm utilizes a four-layer fully connected neural network, with one input layer, two hidden layers with \(256\) neurons, and one output layer. A hyperbolic tangent function is used as the activation function for all layers. The maximum number of training episodes is set to \(10^8\), with each episode containing \(10\) time slots. Following \cite{10233705}, the initial learning rate is set to \(3 \times {10^{ - 4}}\) and decays exponentially as the training episodes progress, with decay steps set to \(10^4\). Additional hyper-parameters of our proposed PPO-based DRL algorithm are listed in Table I.
\vspace{0.5pt}

\begin{table}[h]
    \centering
    \caption{Hyper Parameters Settings}
    \vspace{0.2cm}
    \normalsize
    \renewcommand{\arraystretch}{1.0} 
    \setlength{\tabcolsep}{20pt} 
    \begin{tabular}{|c|c|}
        \hline
        \textbf{Parameter name} & \textbf{Value} \\
        \hline
        Discount factor $\gamma$ & 0 \\
        \hline
        Clipping parameter $\varepsilon$ & 0.2 \\
        \hline
        Memory size & 40960  \\
        \hline
        Batch size & 2048 \\
        \hline
        The number of NN layer & 4 \\
        \hline
    \end{tabular}
\end{table}

\subsection{Results and Discussion}

\begin{figure}[t]
\centering
\includegraphics[scale=0.55]{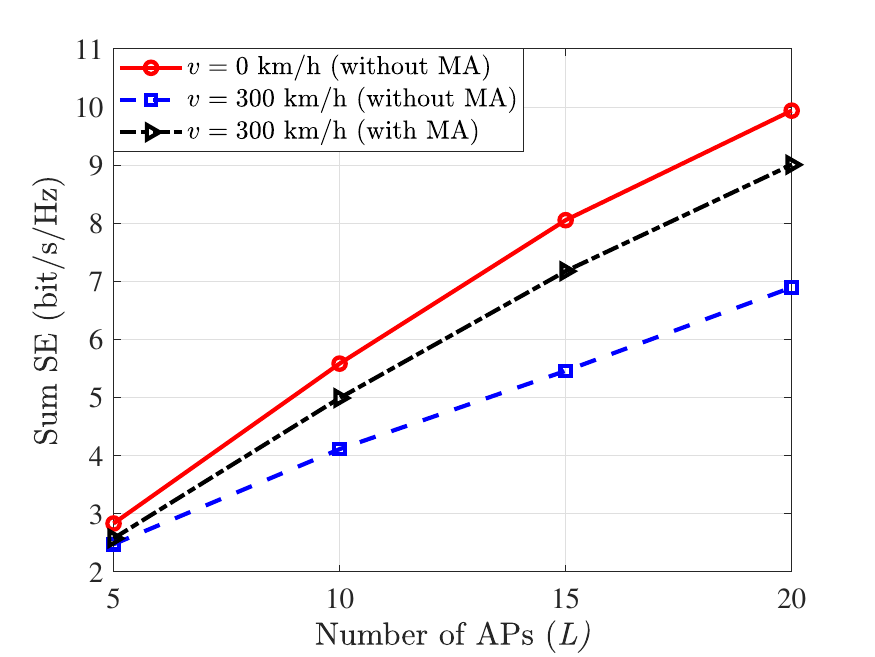}
\caption{Sum SE with the number of APs under different speeds in CF massive MIMO systems ($K\!=\!8$, ${d_s} = \lambda /2$, $d_{\text{ve}}\!=\!50\text{ m}$).} \vspace{-4mm}
\label{figure1}
\end{figure}

\begin{figure}[t]
\centering
\includegraphics[scale=0.55]{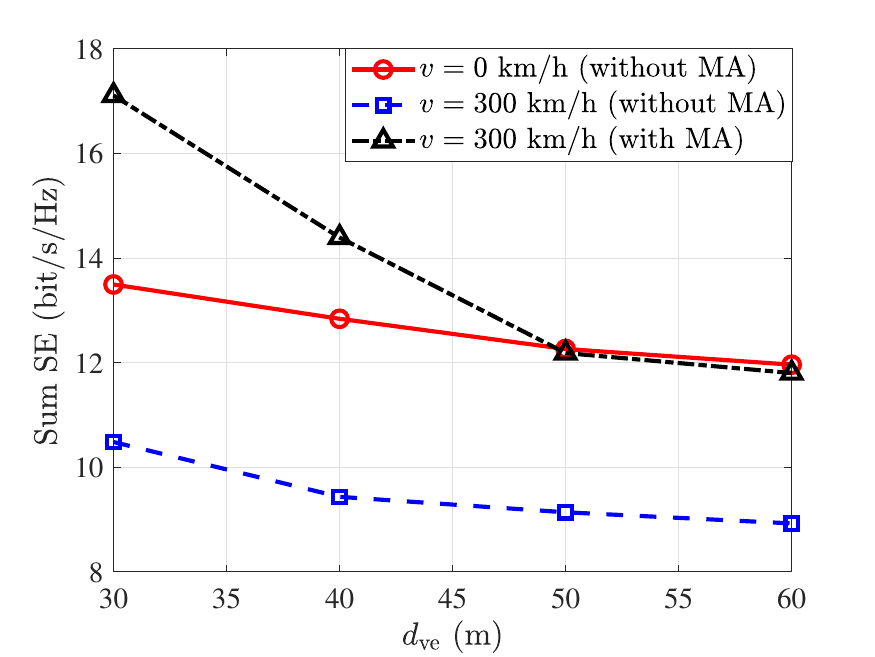}
\caption{Sum SE with the parameter \(d_{\text{ve}}\) under different speeds in CF massive MIMO systems ($K\!=\!8$, $L\!=\!30$, ${d_s} = \lambda /2$).} \vspace{-4mm}
\label{figure2}
\end{figure}

Fig. 2 illustrates the sum SE with the number of APs under different speeds in CF massive MIMO systems with FPAs for HST communications. For a stationary HST, the sum SE increases significantly as the number of APs grows from \(L = 5\) to \(L = 20\). This improvement is due to the CPU’s coordination of distributed APs, providing additional spatial degrees of freedom (DoF) and array gains to mitigate interference \cite{ngo2013energy}. However, the sum SE for a moving HST remains lower and increases less significantly. This is due to phase noise and handover issues that cause inter-carrier interference (ICI) and degrade performance. Fig. 3 shows the impact of the vertical distance \(d_{\text{ve}}\) between APs and TAs on sum SE under different speeds. As the distance increases, sum SE decreases in both cases. Moreover, a moving HST reduces sum SE by approximately 28\% to 34\%. Thus, the PPO-based algorithm is introduced to optimize antenna positions for moving HST, effectively balancing Doppler effects and improving sum SE. The PPO-based algorithm also outperforms in optimizing scenarios with more APs or smaller vertical distance \(d_{\text{ve}}\).

To demonstrate the convergence of the PPO-based algorithm, we plot the training process for the sum SE, smoothed with a sliding window to highlight trends. Fig. 4 shows the sum SE for all TAs versus training episodes. As training progresses, the total SE gradually converges, with diminishing fluctuations. This convergence results from efficient sample use in the experience pool, where the PPO algorithm balances exploration and exploitation for stable learning. The positions of MAs are either uniformly discrete or continuously adjustable. For discrete positions, the spacing interval is either \(d_s=\lambda\) or \(d_s={\lambda}/{2}\). A smaller interval \(d_s\) increases the final sum SE but slows convergence. As \(d_s \to 0\), positions become continuously adjustable, yielding a higher final sum SE.
 
\begin{figure}[t]
\centering
\includegraphics[scale=0.55]{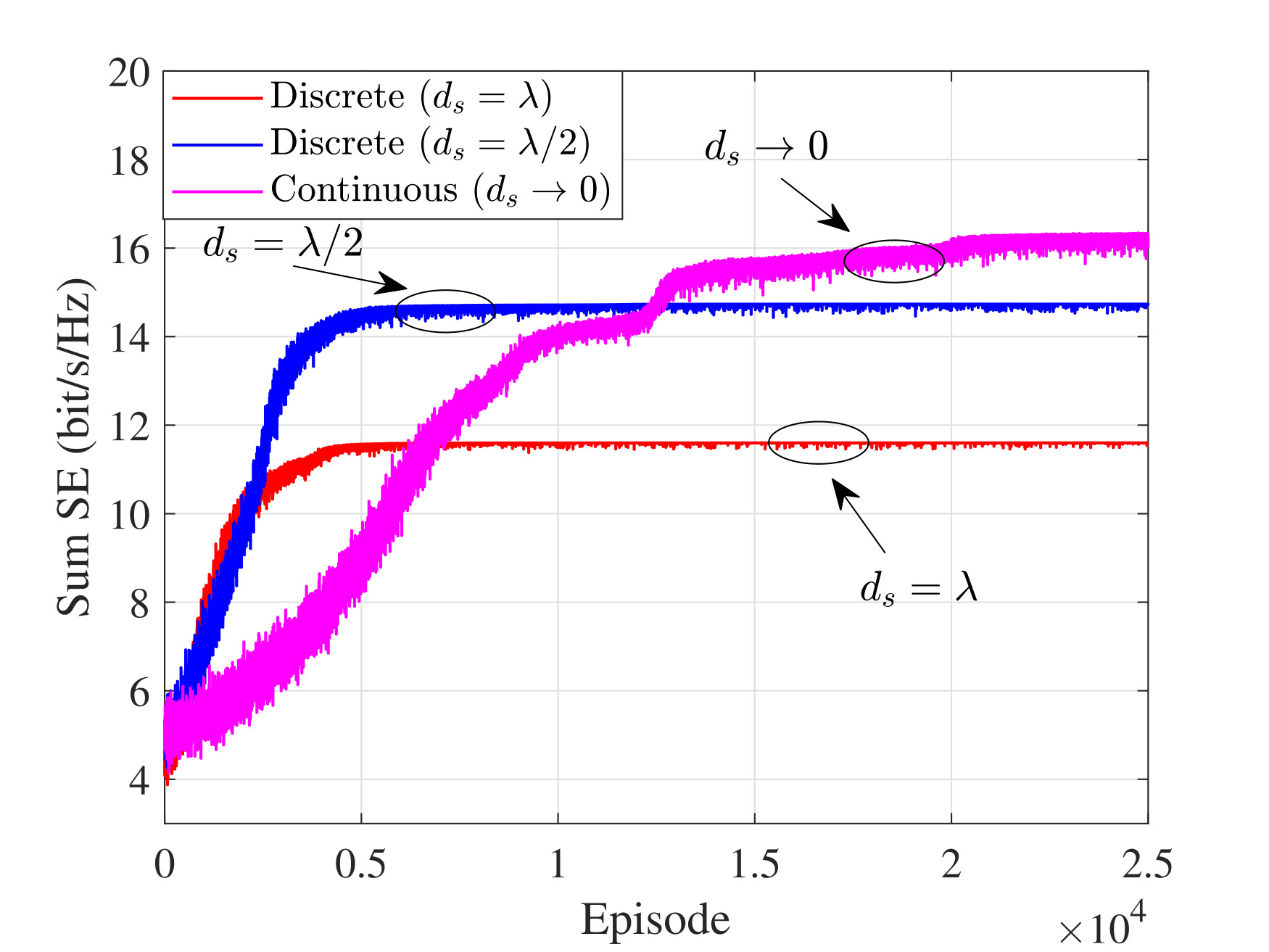}
\caption{Convergence of the PPO-based algorithm ($L\!=\!30$, $K\!=\!8$, $N\!=\!10$, $v\!=\!300\text{ km/h}$, $d_{\text{ve}}\!=\!50\text{ m}$).} 
\vspace{-4mm}
\label{figure4}
\end{figure}

\begin{figure}[t]
\centering
\includegraphics[scale=0.55]{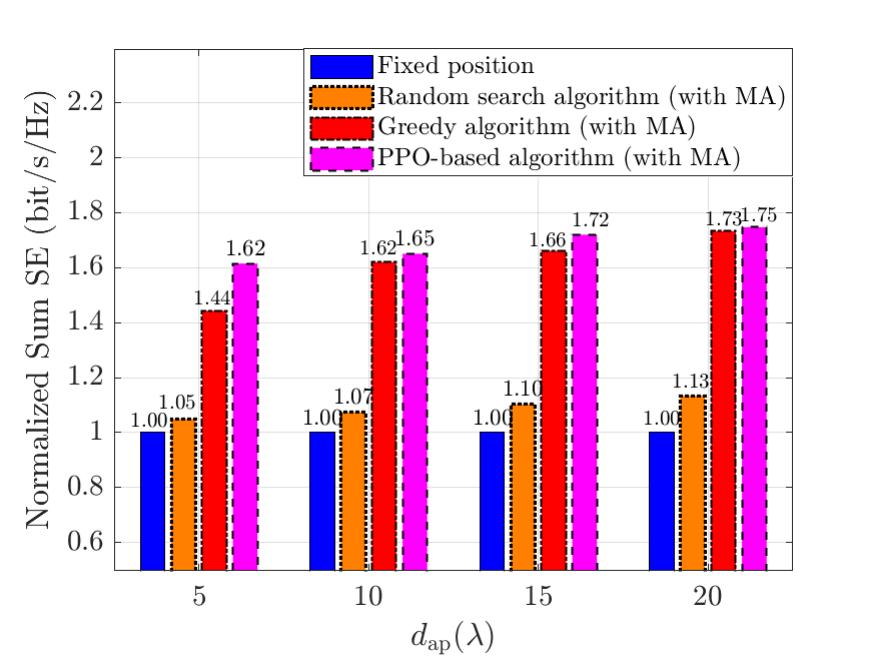}
\caption{Comparison of the PPO-based algorithm with random search and greedy algorithms ($L\!=\!30$, $K\!=\!8$, $v\!=\!300\text{ km/h}$, $d_{\text{ve}}\!=\!50\text{ m}$).} \vspace{-4mm}
\label{figure3}
\end{figure}

In Fig. 5, we compare the effectiveness of antenna positions that are fixed with those optimized by the random search, greedy or PPO-based algorithms for HST communications. The interval of the discrete positions of MA on the AP is \({\lambda }/2\). The results show that for a fixed AP length, the PPO-based algorithm outperforms the other two algorithms with higher sum SE. It is because the greedy algorithm relies on immediate rewards and lacks adaptive learning capabilities \cite{5751501}, while the random search algorithm performs worse due to its lack of strategic exploration. Additionally, as the AP length increases from \(d_{\rm{ap}} = 5\lambda \) to \(d_{\rm{ap}} = 20\lambda \), the sum SE for all algorithms also increases. This is because a longer AP length provides more potential positions for MA, allowing for more reasonable of its position. However, the performance gap between the greedy and PPO-based algorithms diminishes with increasing AP length. This is because more available positions allow both algorithms to explore solution, thereby reducing the relative superiority of the PPO-based algorithm compared to the greedy algorithm.

\section{Conclusion}\label{se:conclusion}
In this paper, we investigated the performance of movable CF massive MIMO for HST communications. Considering the LoS transmission and the DFO effect, the uplink SE expression for the considered system are derived. Additionally, we formulated an optimization problem to maximize the sum SE by optimizing the positions of MAs. Importantly, a PPO-based algorithm is proposed to address this complex problem. The simulation results showed that utilizing the PPO-improved algorithm to optimize the positions of MAs can effectively mitigate the negative impact of DFO, which was demonstrated superior sum SE performance and adaptability in dynamic environments compared to the greedy algorithm.

\vspace{0.3cm}

\bibliographystyle{IEEEtran}
\bibliography{IEEEabrv,Ref}

\end{document}